 \documentclass[final,3p,times]{elsarticle}
\usepackage{amssymb}
\usepackage{amsmath}
\begin{document}

\begin{frontmatter}



\title{Two-cluster regular states, chimeras and hyperchaos in a system of globally coupled phase oscillators with inertia}


\author[inst1]{Vyacheslav O. Munyayev}

\affiliation[inst1]{organization={Department of Control Theory, Scientific and Educational Mathematical Center “Mathematics of Future Technologies”, Nizhny~Novgorod~State~University},%Department and Organization
            addressline={Gagarin~Ave.~23}, 
            city={Nizhny Novgorod},
            postcode={603022}, 
            country={Russia}}

\author[inst1]{Maxim~I.~Bolotov\corref{cor1}}
\ead{maxim.i.bolotov@gmail.com}
\cortext[cor1]{Corresponding author}
\author[inst1]{Lev~A.~Smirnov}
\author[inst1]{Grigory~V.~Osipov}

\begin{abstract}
In this work, two-cluster modes are studied in a system of globally coupled Kuramoto--Sakaguchi phase oscillators with inertia. It is shown that these regimes can be of two types: with a constant intercluster phase difference rotating at the same frequency (according to the analysis, such regimes are always unstable) and with a periodically changing (taking into account the multiplicity of $2\pi$) phase mismatch. The issues of existence and stability, emergence and destruction of two-cluster modes are studied depending on the parameters: effective mass (responsible for inertial processes in the model system under consideration) and phase shift in the coupling function. The analytical results are confirmed and supplemented by numerical simulation of the rotators (second order) interacting globally through the mean field.
\end{abstract}



\begin{keyword}
phase oscillator \sep inertia \sep global coupling \sep cluster \sep chimera state \sep hyperchaos \sep stability analysis
\PACS 05.45.Xt \sep 05.45.-a
\MSC 34C15 \sep 34C28
\end{keyword}

\end{frontmatter}


\section{Introduction}
\label{sec:intr}
Ensembles of phase oscillators are often used to model collective dynamics for the large number of interacting elements. They are one of the paradigmatic models for a fundamental understanding of the behavior and analysis of key processes and modes in various nature systems, ranging from neural networks~\cite{hoppensteadt2012}, populations of chemical oscillators~\cite{tinsley2012} to laser gratings~\cite{ding2019} and electrical networks~\cite{dorfler2013}. The first-order Kuramoto phase oscillator system is a widely adapted and frequently used model of network activity, which, in particular, includes non-trivial cooperative dynamics and self-organization of clusters, and also clearly demonstrates the transition from incoherent to synchronous modes states~\cite{acebron2005,barreto2008,ott2008,hong2007,pikovsky2008,maistrenko2004,dorfler2011,martens2009}. Note that when the natural frequencies of the element rotations are non-uniform, then the transition to partial phase synchronization is characteristic. In this case, the system breaks up into clusters of coherent and incoherent oscillators~\cite{acebron2005,martens2009,laing2009}.

The second-order Kuramoto model with inertia is commonly used to describe networks of oscillators capable of regulating their own frequencies, as in the adaptive frequency timing model of fireflies~\cite{ermentrout1991} and power grid systems~\cite{tumash2019}. The inclusion of inertia leads to two-dimensional intrinsic dynamics of the oscillators, thereby making the collective dynamics of the second-order Kuramoto model significantly more complex compared to the dynamics of the classical first-order model. This dynamics, associated with taking into account inertia, includes complex transitions from incoherence to complete synchronization, bistability of synchronous clusters, bistability of synchronous clusters, chaotic intercluster dynamics \cite{brister2020}, chimeras~\cite{olmi2015,maistrenko2017,medvedev2021}, solitary states~\cite{jaros2015,jaros2018,munyayev2022}, as well as cyclops regimes~\cite{munyayev2023}.

In the article~\cite{belykh2016} the emergence and coexistence of stable clusters in a two-population network of identical second-order Kuramoto oscillators was analytically studied. Two populations of different sizes $K$ and $M$ naturally separate into two clusters, within which the oscillators synchronize, creating a phase shift between the clusters. The analysis performed in~\cite{belykh2016} made it possible to obtain the necessary and sufficient conditions for the stability of two-cluster synchronization, characterized by a constant phase difference between the elements of the clusters, and also provided a stability condition for two-cluster synchronization with a rotating phase difference.

In the work~\cite{munyayev2022}, the so-called solitary states were studied; they are characterized by the elements divided into two clusters, one of which contains only one element. This state is called solitary. The article~\cite{munyayev2023} described an amazing three-cluster regime -- the cyclops state. The cyclops mode can be realized in ensembles of phase oscillators with inertia, containing an odd number of elements, with the elements breaking up into two equal-sized clusters and one solitary element. The purpose of this work is to study two-cluster modes in the Kuramoto--Sakaguchi system with inertia.

Chapter~\ref{sec:I} discusses the model of globally coupled phase oscillators with inertia and its parameters. A description of the single-cluster and two-cluster rotational modes is given, and their connection with the dynamics of a dissipative pendulum under the influence of a constant torque is presented. Chapter~\ref{sec:II} analyzes the existence of two-cluster regimes with a constant and periodically rotating phase mismatch between the clusters. Chapter~\ref{sec:III} describes a linear analysis of the stability of two-cluster modes, obtains variational equations, and describes the modes responsible for the tangential and transversal stability of clusters. In Chapter~\ref{sec:IV}, based on direct numerical modeling of a system of phase oscillators with inertia, scenarios for the destruction of two-cluster states and the resulting regimes are demonstrated. Finally, in Chapter~\ref{sec:con} the main results of the work are formulated.

\section{The Kuramoto model with inertia. Single-cluster and two-cluster modes}\label{sec:I}
\subsection{Model of phase oscillators with inertia. Full synchronization}
Let us consider a model of $N$ globally coupled Kuramoto--Sakaguchi phase oscillators with inertia (index $n = 1, 2, \dots, N$), described by the following equations:
\begin{equation}
  m\ddot{\varphi}_n + \dot{\varphi}_n = \omega + \frac{1}{N} \sum_{\tilde{n} = 1}^N \sin{\left(\varphi_{\tilde{n}} - \varphi_n - \alpha\right)}, 
\label{eq:main-system}
\end{equation}
where $\varphi_n(t)$ are the phases of the oscillators, $m$ is the inertia parameter (or mass), $\omega$ is the frequency (or torque), $\alpha$ is the phase shift (Sakaguchi parameter). The oscillators are assumed to be identical with mass $m$, frequency $\omega$ and phase shift $\alpha \in [0,\pi]$~\cite{sakaguchi2006}. Due to the symmetry, the system~\eqref{eq:main-system} has a global in-phase rotational mode
\begin{equation}
  \varphi_1\left(t\right)=\varphi_2\left(t\right)=\ldots=\varphi_N\left(t\right),
\label{eq:globs}
\end{equation}
which is locally stable for any $\alpha\in\left[0,\pi/2\right)$ and unstable for any $\alpha\in\left(\pi/2,\pi\right]$~\cite{acebron2005} Therefore, it is customary to consider the connection to be attractive in the first case and repulsive in the second case.

\subsection{Two-cluster synchronous modes. Basic equations}
In addition to the global synchronization mode~\eqref{eq:globs}, in systems of globally coupled elements, cluster synchronous modes can occur, in which the entire ensemble can be divided into groups -- clusters of mutually synchronous elements. In this case, each cluster has its own oscillation frequency, different from other frequencies (for the system considered here -- rotations). This situation can occur in ensembles of non-identical elements, when frequency synchronization is of interest. In the case of identical elements, as in this problem, phase synchronization is also interesting. With cluster phase synchronization, in each cluster all elements perform in-phase oscillations (rotations). And between elements from different clusters there is either a constant or time-invariant phase shift. It is obvious that with a non-constant phase shift, the rotation frequencies in clusters are generally different. Note that at some parameter values synchronization of rotations can be observed between the clusters. We consider the two-cluster mode of phase synchronization in the model~\eqref{eq:main-system}. For definiteness, we assume that the small synchronous cluster contains $K$ elements. Then
\begin{equation}
\begin{gathered}
  \varphi_1\left(t\right)=\varphi_2\left(t\right)=\ldots=\varphi_K\left(t\right)=\psi_{1}\left(t\right), \\
  \varphi_{K+1}\left(t\right)=\varphi_{K+2}\left(t\right)=\ldots=\varphi_N\left(t\right)=\psi_{2}\left(t\right),
\end{gathered}
\label{eq:two-cluster-def}
\end{equation}
which leads to the following system of equations:
\begin{equation}
\begin{gathered}
  m\ddot{\psi}_1 + \dot{\psi}_1 = \omega - \frac{K}{N}\sin{\alpha} + \frac{N-K}{N} \sin{\left(\psi_2 - \psi_1 - \alpha\right)}, \\
  m\ddot{\psi}_2 + \dot{\psi}_2 = \omega - \frac{N-K}{N}\sin{\alpha} + \frac{K}{N} \sin{\left(\psi_1 - \psi_2 - \alpha\right)},
\end{gathered}
\label{eq:two-cluster}
\end{equation}
where $\psi_1$, $\psi_2$ are the first and second cluster phases, respectively, $K$ is the number of elements in the small cluster. Let us define a new parameter $\beta = {K}/{N}\le0.5$, which characterizes the proportion of elements of a small cluster relative to the size $N$ of the entire ensemble. Then the system~\eqref{eq:two-cluster} can be rewritten as:
\begin{equation}
\begin{gathered}
  m\ddot{\psi}_1 + \dot{\psi}_1 = \omega - \beta\sin{\alpha} + \left(1 - \beta\right) \sin{\left(\psi_2 - \psi_1 - \alpha\right)}, \\
  m\ddot{\psi}_2 + \dot{\psi}_2 = \omega - \left(1 - \beta\right)\sin{\alpha} + \beta \sin{\left(\psi_1 - \psi_2 - \alpha\right)}.
\end{gathered}
\label{eq:two-cluster-beta}
\end{equation}

Let us introduce a new variable $\theta(t)=\psi_1(t)-\psi_2(t)$, characterizing the intercluster phase difference. Subtracting the second one from the first equation in~\eqref{eq:two-cluster-beta}, we arrive at the equation for $\theta(t)$:
\begin{equation}
  m\ddot{\theta} + \dot{\theta} = \left(1 - 2 \beta\right) \sin{\alpha} - \left[\left(1-\beta\right)\sin{\left(\theta + \alpha\right)} + \beta\sin{\left(\theta - \alpha\right)} \right].
\label{eq:pre-pend}
\end{equation}
Note that the solution $\theta_1 = 0$ corresponds to the in-phase rotational mode~\eqref{eq:globs}. It exists for any $\alpha$ and $\beta$. The non-zero solution $\theta_2=\mathrm{const}\neq0$ corresponds to a two-cluster regime with a constant phase mismatch in the original system~\eqref{eq:main-system}. A periodic rotational solution $\theta_r(t)$ satisfying the condition $\theta_r\!\left(t+T_0\right)\equiv\theta_r\!\left(t\right)\pmod{2\pi}$ corresponds to a two-cluster regime with a periodically changing intercluster phase difference in the original system~\eqref{eq:main-system}. Here $T_0$ is the time during which the variable $\theta$ changes by $2\pi$ (period). In what follows, we call such a regime for the original system a rotational two-cluster regime. Other two-cluster modes, for example oscillatory, for which $\left|\theta\!\left(t\right)\right|\le\mathrm{const}$, do not exist in the system~\eqref{eq:main-system}.
Eq.~\eqref{eq:pre-pend} is equivalent to the equation of a pendulum with dissipation and constant external torque. In fact, replacement
\begin{equation}
  \theta\!\left(t\right)=\Phi\left(\sqrt{{R}/{m}}\,t\right)-\delta
\end{equation}
leads to the pendulum equation
\begin{equation}
  \Phi''+\gamma\Phi'+\sin\Phi=\tau,
\label{eq:pend}
\end{equation}
where $R\!=\!\sqrt{\left(1-2\beta\right)^2\sin^2\!{\alpha}+\cos^2\!{\alpha}}$, $\delta\!=\!\arccos\!\left({\cos\alpha}/{R}\right)$, and the parameters $\gamma\!=(mR)^{-1/2}$, $\tau\!=\!(1-2\beta)\sin\alpha/R$.

The resulting pendulum equation~\eqref{eq:pend} has been well studied (see, for example,~\cite{andronov2013}). Depending on the parameters $\gamma$ and $\tau$, two equilibrium states can exist in the system~\eqref{eq:pend}: stable ($\Phi_1 = \arcsin{\tau}$, $\Phi'_1 = 0$) and saddle ($\Phi_2 = \pi - \arcsin{\tau}$, $\Phi'_2 = 0$) (exist for $\tau < 1$), as well as stable periodic rotational motion $\Phi_r(t)$. It exists in the region determined approximately by the inequality $\tau>\mathcal{T}\left(\gamma\right) \approx 4\gamma/\pi-0.305\gamma^3$~\cite{belykh2016}, where $\gamma<\gamma^{*}\approx{1.1865}$ and $\tau>1$ if $\gamma>\gamma^{*}$, where $\mathcal{T}\left(\gamma\right )$ -- Tricomi bifurcation curve. The presence of two attractors (stable equilibrium $\Phi=\Phi_1$ and stable periodic rotational motion $\Phi(t) = \Phi_r(t)$) leads to the fact that two types of two-cluster rotational regimes can exist in the original system~\eqref{eq:main-system}. The first is characterized by a constant phase mismatch $\Phi_1$, the second is characterized by a periodically increasing phase mismatch $\Phi_r\left(t\right)$.

In the case of periodic phase mismatch $\theta_r\!\left(t\right)$, according to the system~\eqref{eq:two-cluster-beta}, the average frequencies of the small and large cluster are equal to $\Omega_1$ and $\Omega_2$, respectively:
\begin{equation}
\begin{gathered}
\Omega_1=\omega-\beta\sin{\alpha}-\frac{1-\beta}{T_0}\int\limits_0^{T_0}{\sin\left(\theta_r\!\left(t\right)+\alpha\right)\mathrm{d}t}, \\
\Omega_2=\omega-\left(1-\beta\right)\sin{\alpha}+\frac{\beta}{T_0}\int\limits_0^{T_0}{\sin\left(\theta_r\!\left(t\right)-\alpha\right)\mathrm{d}t}.
\end{gathered}
\label{eq:Omega}
\end{equation}
When $\theta_r\!\left(t\right)$ is linear growing (decreasing), we obtain a simple approximation for frequencies $\Omega_1$, $\Omega_2$:
\begin{equation}
  \Omega_1\approx\omega-\frac{K}{N}\sin\alpha,\quad
  \Omega_2\approx\omega-\frac{N-K}{N}\sin\alpha.
\label{eq:Omega_approx}
\end{equation}

\section{Two-cluster regimes existence}
\label{sec:II}

\subsection{Two-cluster regime with constant intercluster phase difference}
It should be noted that the parameterization used to derive Eq.~\eqref{eq:pend} always satisfies the inequality $0\le\tau\le1$. Thus, in the original system~\eqref{eq:pre-pend} for any parameters there are always two equilibrium states. Finding them is not difficult:
\begin{equation}
\begin{gathered}
  \theta_1=0,\quad\dot{\theta}_1=0, \\
  \theta_2=2\arctan\!\big({\cot\alpha}/{(1-2\beta)}\big),\quad \dot{\theta}_2=0.
\end{gathered}
\end{equation}
Analysis of linear stability shows that the state $\theta_2$, corresponding to two-cluster dynamics with a constant phase mismatch, is unstable (saddle) for $0<\alpha<\pi/2$. In this case, the state $\left(\theta_1,0\right)$, corresponding to global synchronization, is stable. At the point $\alpha=\pi/2$ a transcritical bifurcation is observed, as a result of which the states $\left(\theta_1,0\right)$ and $\left(\theta_2,0\right)$ exchange stability. Thus, for $\pi/2<\alpha<\pi$ the state $\left(\theta_2,0\right)$ is stable, and the state $\left(\theta_1,0\right)$ is unstable (in-phase mode is unstable).

\subsection{Two-cluster regime with rotating intercluster phase difference}
\begin{figure}
	\begin{center}
        \includegraphics[width=1\columnwidth]{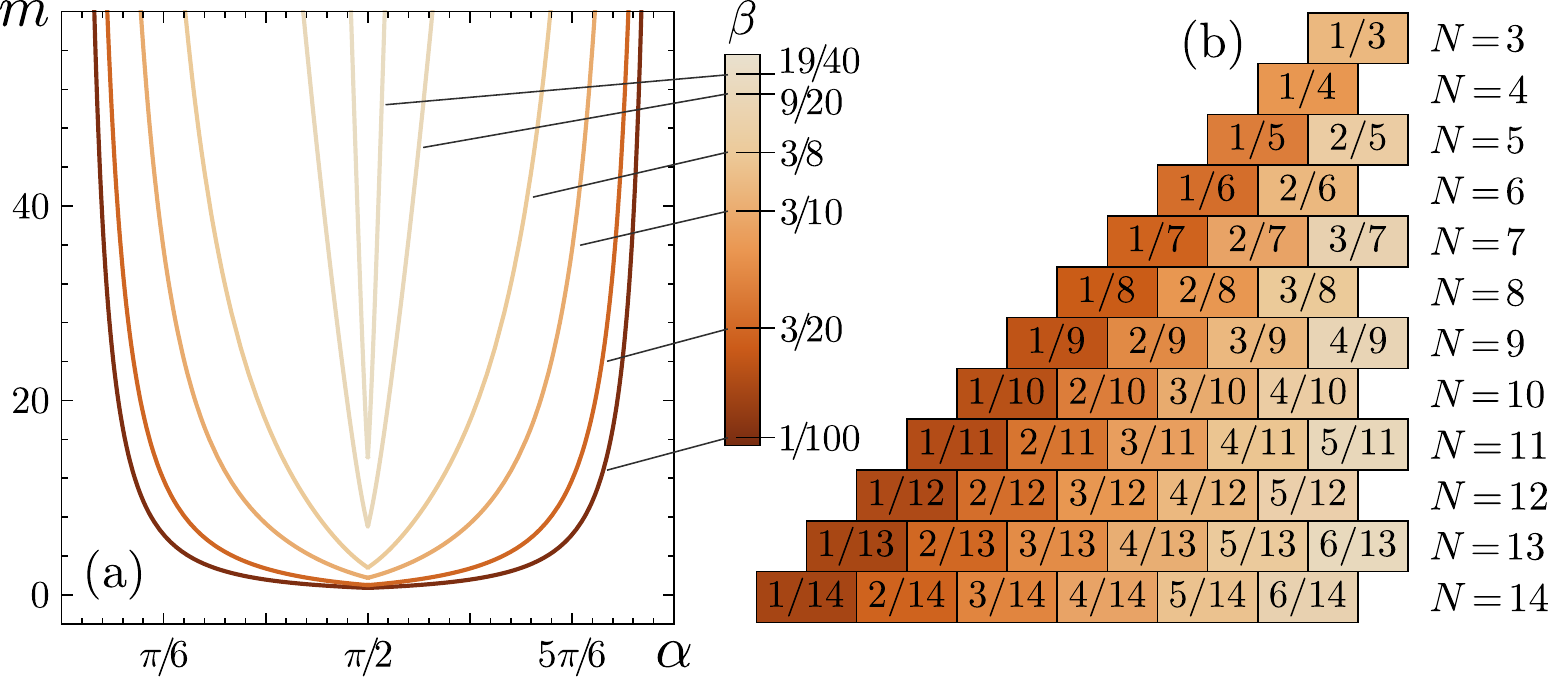}
	\end{center}
	\caption{(a) Boundaries of the two-cluster rotational mode~\eqref{eq:two-cluster-def} regions existence with periodic rotating intercluster phase difference $\theta_r\!\left(t\right)$, on the parameter plane $(\alpha, m) $ for different $\beta$ (color-coded). For fixed $\beta$, the regime exists in the region defined by the inequality~\eqref{eq:eroi}. (b) Diagram of two-cluster modes~\eqref{eq:two-cluster-def} with rotating phase mismatch $\theta_r\!\left(t\right)$ depending on the number of elements $N$. Each cell corresponds to a specific two-cluster regime, characterized by the parameter $\beta = K/N$, which is written as a fraction in the center of the cell. Cells with the same color correspond to two-cluster rotational modes with the same $\beta$ values.}
	\label{fig1}
\end{figure}

Since in the original problem the inequality $0\le\tau\le1$ is always satisfied automatically, the region of existence of stable periodic rotational motion in the system~\eqref{eq:pend} is determined only by the inequality
\begin{equation}
	\tau > \mathcal{T}\left(\gamma\right),
\end{equation}
where $\gamma\le\gamma^{*}$. In the initial parameters of the problem, the inequality takes the form:
\begin{equation}
  {(1-2\beta)}\sin{\alpha}/R > \mathcal{T}\left((mR)^{-1/2}\right).
\end{equation}
In the general case, taking into account the monotonicity of the function $\mathcal{T}\left(\gamma\right)$, we find the inequality that determines the region of existence of two-cluster rotational motion with periodic phase mismatch on the parameter plane $\left(\alpha,m\right)$:
\begin{equation}
  m\!>\!\frac{\left[\mathcal{T}^{-1}\big({\left(1-2\beta\right)\sin{\alpha}}\big({\left(1-2\beta\right)^2\sin^2\!{\alpha}+\cos^2\!{\alpha}})^{-1/2}\big)\right]^{-2}}{\big({\left(1-2\beta\right)^2\sin^2\!{\alpha}+\cos^2\!{\alpha}}\big)^{1/2}}.
\label{eq:eroi}
\end{equation}
%\begin{equation}
%  m\!>\!\frac{1}{\sqrt{\left(1-2\beta\right)^2\sin^2\!{\alpha}+\cos^2\!{\alpha}}}\!\left[\mathcal{T}^{-1}\!\left(\frac{\left(1-2\beta\right)\sin{\alpha}}{\sqrt{\left(1-2\beta\right)^2\sin^2\!{\alpha}+\cos^2\!{\alpha}}}\right)\right]^{-2}\!.
%\label{eq:eroi}
%\end{equation}
Fig.~\ref{fig1}(a) shows the boundaries of the regions of existence of two-cluster rotational modes with rotating phase mismatch for various values of the parameter $\beta$. It can be noted that the two-cluster mode existence region, characterized by the parameter $\beta_1$, is nested in the region of existence of the two-cluster mode with the parameter $\beta_2$ at $\beta_1>\beta_2$. The coexistence of rotational modes is observed. The expression~\eqref{eq:eroi} allows us to find the minimum value of $m$ at which the existence of the considered modes is possible. This value is achieved at $\alpha=\pi/2$ and is equal to
\begin{equation}
  m^{*}={{\gamma^{*}}^{-2}\left({1-2\beta}\right)^{-1}}.
\end{equation}

From this expression it follows that the two-cluster mode with rotational mismatch at $\beta=0.5$, i.e. when the entire ensemble is divided into two clusters of equal size does not exist (since $m^{*}\to\infty$).

Fig.~\ref{fig1}(b) shows a diagram (in the form of a pyramid) of two-cluster rotational modes depending on the parameter $N$. Each cell corresponds to a specific two-cluster regime, characterized by the parameter $\beta$, which is written as a fraction in the center of the cell. Cells with the same color correspond to two-cluster rotational modes with the same $\beta$ value. It is easy to see that for a fixed $N = N^*$, the two-cluster rotational mode corresponding to $\beta^* = A/B$, where $A/B$ is some fraction, will be repeated for all $N = N^* + k\cdot B$ ($k$ is an arbitrary integer). Using this representation and combinatorial considerations, it is easy to determine the number of two-cluster partitions for a given number of elements $N$. The required number of two-cluster modes is equal to $\lfloor(N-1)/2\rfloor$ ($\lfloor\ldots\rfloor$ -- rounding down).

Thus, most two-cluster modes coexist in the high mass limit, that is called the underdamped limit. At relatively small masses, there are only strongly unbalanced two-cluster partitions (in particular, solitary states), which correspond to the left slanted edge of the diagram.

\section{Stability of two-cluster regimes}
\label{sec:III}
\subsection{Equations in variations. Tangential and transversal stability}

To perform the stability analysis of two-cluster regimes, we start with the general equations for variations $\delta\varphi_n$ of the system~\eqref{eq:main-system}:
\begin{equation}
  m\delta{\ddot\varphi_n} + \delta{\dot\varphi_n} = \frac{1}{N}\sum\limits_{\tilde{n} = 1}^N{\cos{\left(\varphi_{\tilde{n}} - \varphi_n - \alpha\right)}\left(\delta\varphi_{\tilde{n}} - \delta\varphi_n\right)}.
  \label{eq:variations}
\end{equation}
Given~\eqref{eq:two-cluster-def}, for the two-cluster regimes we find that
\begin{equation}
\begin{gathered}
  m\delta{\ddot\varphi_n} + \delta{\dot\varphi_n} = \frac{\cos\alpha}{N}\sum\limits_{{\tilde{n}} = 1}^K{\left(\delta\varphi_{\tilde{n}} - \delta\varphi_n\right)} + \frac{\cos{\left(\theta\!\left(t\right)+\alpha\right)}}{N}\sum\limits_{{\tilde{n}} = K+1}^N{\left(\delta\varphi_{\tilde{n}} - \delta\varphi_n\right)}, \hspace{0.2cm} 1\le{n}\le{K},\\
  m\delta{\ddot\varphi_n} + \delta{\dot\varphi_n} = \frac{\cos{\left(\theta\!\left(t\right)-\alpha\right)}}{N}\sum\limits_{{\tilde{n}} = 1}^K{\left(\delta\varphi_{\tilde{n}} - \delta\varphi_n\right)} + \frac{\cos\alpha}{N}\sum\limits_{{\tilde{n}} = K+1}^N{\left(\delta\varphi_{\tilde{n}} - \delta\varphi_n\right)}, \hspace{0.2cm} K<{n}\le{N}.
\end{gathered}
\end{equation}
For further analysis we substitute the variables
\begin{align}
  \eta_1 &= \frac{1}{K} \sum_{{\tilde{n}} = 1}^K \delta\varphi_{\tilde{n}} - \frac{1}{N - K} \sum_{{\tilde{n}} = K + 1}^N \delta\varphi_{\tilde{n}}, \\
  \eta_2 &= \frac{1}{K} \sum_{{\tilde{n}} = 1}^K \delta\varphi_{\tilde{n}} + \frac{1}{N - K} \sum_{{\tilde{n}} = K + 1}^N \delta\varphi_{\tilde{n}}, \\
  \xi_n &= \delta\varphi_{n+1} - \delta\varphi_n, \hspace{1.5cm} 1 \leq n < K, \\
  \zeta_n &= \delta\varphi_{n+1} - \delta\varphi_n, \hspace{1.5cm} K < n < N.
\end{align}
We arrive at a system of equations for tangential stability of the set of two-cluster solutions~\eqref{eq:two-cluster-def}
\begin{equation}
\begin{gathered}
  m\ddot{\eta}_1 + \dot{\eta}_1 + \left[ \beta \cos{(\theta\!\left(t\right) - \alpha)} + (1 - \beta) \cos{(\theta\!\left(t\right) + \alpha)} \right] \eta_1 = 0, \\
  m\ddot{\eta}_2 + \dot{\eta}_2 + \left[ (1 - \beta) \cos{(\theta\!\left(t\right) + \alpha)} - \beta \cos{(\theta\!\left(t\right) - \alpha)} \right] \eta_1 = 0,
\end{gathered}
\label{eq:split-linear-pert-sys-n12}
\end{equation}
and transversal stability
\begin{equation}
\begin{gathered}
  m\ddot{\xi}_n + \dot{\xi}_n + \left[ (1 - \beta) \cos{(\theta\!\left(t\right) + \alpha)} + \beta \cos{\alpha} \right] \xi_n = 0, \hspace{0.3cm} 1 \leq n < K, \\
  m\ddot{\zeta}_n + \dot{\zeta}_n + \left[ (1 - \beta) \cos{\alpha} + \beta \cos{(\theta\!\left(t\right) - \alpha)} \right] \zeta_n = 0, \hspace{0.3cm} K < n < N.
\end{gathered}
\label{eq:split-linear-pert-sys-ksi-eta}
\end{equation}
Here, the first equation for $\xi_n$ describes the transversal stability of the small cluster and the second equation for $\zeta_n$ -- the large cluster. Thus, in general, to analyze the stability of two-cluster regimes, only a system of two second-order ODEs~\eqref{eq:split-linear-pert-sys-n12} and two independent second-order equations~\eqref{eq:split-linear-pert-sys-ksi-eta} (due to their identity for different $n$) need to be investigated. In the case of the solitary state ($K=1$), the equations for $\xi_n$ are not considered.

\subsection{Two-cluster regime with constant intercluster phase difference}

To investigate the stability of the two-cluster regime with constant non-zero intercluster phase difference $\theta_2$ within the original model~\eqref{eq:main-system}, we will use the equations for tangential and transversal stability~\eqref{eq:split-linear-pert-sys-n12} and~\eqref{eq:split-linear-pert-sys-ksi-eta} putting $\theta\! \left(t\right)\equiv{\theta}_2$, which leads to the following system:
\begin{equation}
\begin{gathered}
  m\ddot{\eta}_1 + \dot{\eta}_1 - \cos{\alpha}\,\eta_1 = 0, \\
  m\ddot{\eta}_2 + \dot{\eta}_2 - \cos{\alpha}\left(1-2\beta\right)\frac{1+4\beta\left(1-\beta\right)\sin^2\!{\alpha}}{1-4\beta\left(1-\beta\right)\sin^2\!{\alpha}} \eta_1 = 0, \\
  m\ddot{\xi} + \dot{\xi} - \frac{\left(1-2\beta\right)\cos{\alpha}}{1-4\beta\left(1-\beta\right)\sin^2\!{\alpha}}\xi = 0, \quad
  m\ddot{\zeta} + \dot{\zeta} + \frac{\left(1-2\beta\right)\cos{\alpha}}{1-4\beta\left(1-\beta\right)\sin^2\!{\alpha}}\zeta = 0.
\end{gathered}
\end{equation}
From the characteristic equation for $\eta_1$ and $\eta_2$, which has the form
\begin{equation}
  \lambda\!\left(\lambda+\frac{1}{m}\right)\!\left(\lambda^2+\frac{1}{m}\lambda-\frac{\cos\alpha}{m}\right)=0,
\end{equation}
it follows that the two-cluster regime is tangentially stable only with $\cos{\alpha}<0$, i.e., $\pi/2<\alpha<\pi$. It follows in particular that at $N=2$ the two-cluster mode with constant phase difference $\theta$, which, as can be easily seen from the expression for $\theta_2$, is antiphase, will be stable at $\pi/2<\alpha<\pi$. In the case of the solitary state ($K=1$) at $N\ge3$, the large cluster is transversally stable at $0<\alpha<\pi/2$. In all other cases simultaneous transversal stability of both clusters is impossible: at $0<\alpha<\pi/2$ the large cluster is transversally stable, the small cluster is transversally unstable; at $\pi/2<\alpha<\pi$ the opposite is true.

Thus, at $N\ge3$ any two-cluster regime with constant intercluster phase difference is unstable at any values of the system parameters.

\subsection{Two-cluster regime with periodic rotatory intercluster phase difference}
In Sec.~\ref{sec:II}, it was shown that in the existence region of two-cluster regimes with periodic intercluster phase difference, the rotatory dynamics of the upset on the manifold~\eqref{eq:two-cluster-def}, defined by Eq.~\eqref{eq:two-cluster-beta}, is always stable.

To analyze the stability of the two-cluster regime with periodic rotatory phase difference $\theta_r\!\left(t\right)$ (for which $\theta_r\!\left(t+T_0\right)\equiv\theta_r\! \left(t\right)\pmod{2\pi}$, where $T_0$ is the rotation period) within the original distributed model~\eqref{eq:main-system} we will use Eqs.~\eqref{eq:split-linear-pert-sys-n12}--\eqref{eq:split-linear-pert-sys-ksi-eta}. Since at the considered periodic rotatory dynamics of the $\theta_r\!\left(t\right)$, the right-hand sides of Eqs.~\eqref{eq:split-linear-pert-sys-n12}--\eqref{eq:split-linear-pert-sys-ksi-eta} are periodic in $t$, we can apply Floquet theory and find the multipliers determining the asymptotic behavior of their solutions.
	
We analyze the tangential stability using a system~\eqref{eq:split-linear-pert-sys-n12}. One solution to the equation for $\eta_1$ is the function $\dot{\theta}_r\left(t\right)$, which is easily verified by differentiating the expression~\eqref{eq:pre-pend}. Since $\dot{\theta}_r\left(t\right)$ is periodic, one of the multipliers of the equation for $\eta_1$ is 1. According to the Liouville--Ostrogradsky formula, the second multiplier of this equation is $\exp\!{\left(-T_{0}/m\right)}$. Furthermore, the complete system of equations~\eqref{eq:split-linear-pert-sys-n12} has solution $\eta_1 = 0$, $\eta_2 = \mathrm{const}$, whence it follows that the other of the multipliers is equal to 1. Applying the Liouville--Ostrogradsky formula to the whole system, we find the fourth multiplier, also equal to $\exp\!{\left(-T_0/m\right)}$. Thus, the regime under consideration is always tangentially stable, which agrees with the conclusions drawn when considering the dynamics on the manifold~\eqref{eq:two-cluster-def}.
	
To analyze transversal stability Eqs.~\eqref{eq:split-linear-pert-sys-ksi-eta} are used. Due to the fact that these equations are independent of each other, we can see how the stability of the two-cluster regime with periodic rotatory phase difference changes for each cluster. Depending on the solution $\theta_r\!\left(t\right)$, which in the general case can be found only numerically, the stability can disappear or appear only at one or at two clusters, so we can understand how the two-cluster regime will collapse in case of stability loss.

\section{Scenarios of instability development of two-cluster regimes}
\label{sec:IV}
\begin{figure}
\begin{center}
  \includegraphics[width=1\columnwidth]{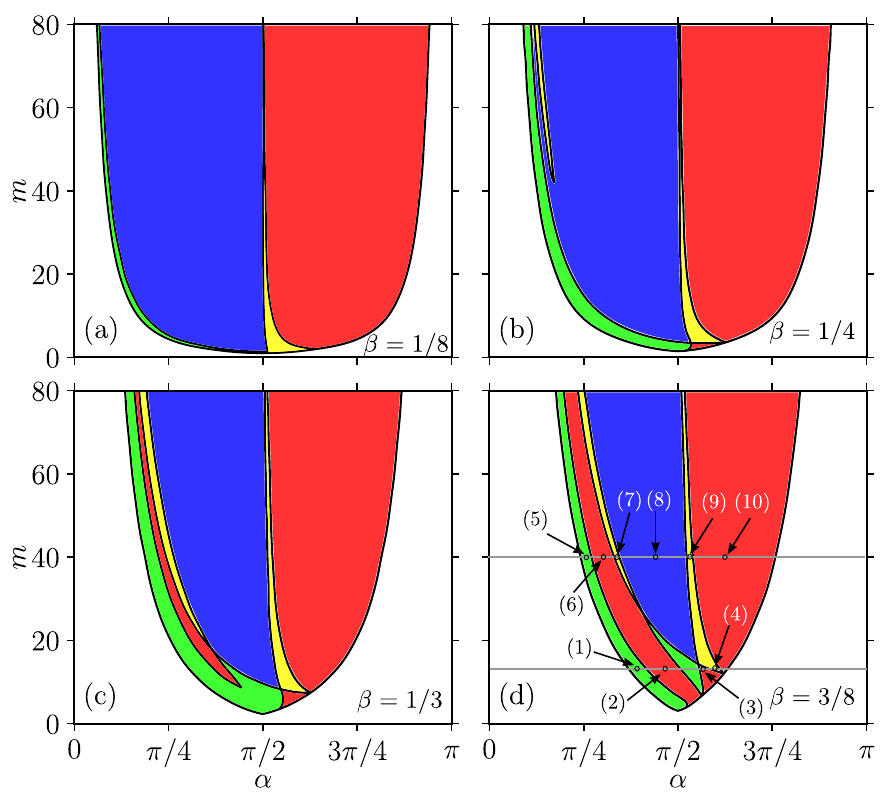}
\end{center}
  \caption{Stability map of two-cluster rotational states~\eqref{eq:two-cluster-def} with rotating intercluster phase difference $\theta_r\!\left(t\right)$ on the parameter plane $(\alpha, m)$. Blue area -- both clusters are stable; green area -- the large cluster is stable, the small one is unstable; yellow area -- the large cluster is unstable, the small one is stable; red area -- both clusters are unstable; white area - mode does not exist. (a) $\beta = 1/8$, (b) $\beta = 1/4$, (c) $\beta = 1/3$, (d) $\beta = 3/8$. The numbered points on the fragment (d) indicate the parameter values for which Fig.~\ref{fig4}--Fig.~\ref{fig8} presents the results of direct numerical modeling of the system~\eqref{eq:main-system}.}
\label{fig2}
\end{figure}

This section analyzes the evolution of two-cluster regimes when the control parameters are changed. As a result of the above analysis of the equations for tangential (Eq.~\eqref{eq:split-linear-pert-sys-n12}) and transversal (Eq.~\eqref{eq:split-linear-pert-sys-ksi-eta}) stability at several fixed values of the parameter $\beta$, maps of stability (instability) regions of two-cluster regimes with periodic rotatory phase difference were obtained. They are represented in the $\left(\alpha,m\right)$ parameter plane (Fig.~\ref{fig2}). The Fig.~\ref{fig2} highlights several main areas: (a) blue area -- both clusters are stable; (b) green area -- large cluster is stable, small cluster is unstable; (c) yellow area -- large cluster is unstable, small cluster is stable; (d) red area -- both clusters are unstable; (f) white area -- the considered two-cluster regime does not exist. It is important to note that the obtained analytical stability (instability) conditions of the two-cluster rotatory regime guarantee its stability (instability) within the original system of equations. At the same time, analytical conditions for the stability of one of the clusters do not guarantee the stability of this cluster within the original model. This is due to the fact that a stable cluster is exposed to the impact of an unstable cluster and under certain conditions this impact can be critical: a stable cluster becomes unstable and collapses.

\begin{figure}
\begin{center}
  \includegraphics[width=1.0\columnwidth]{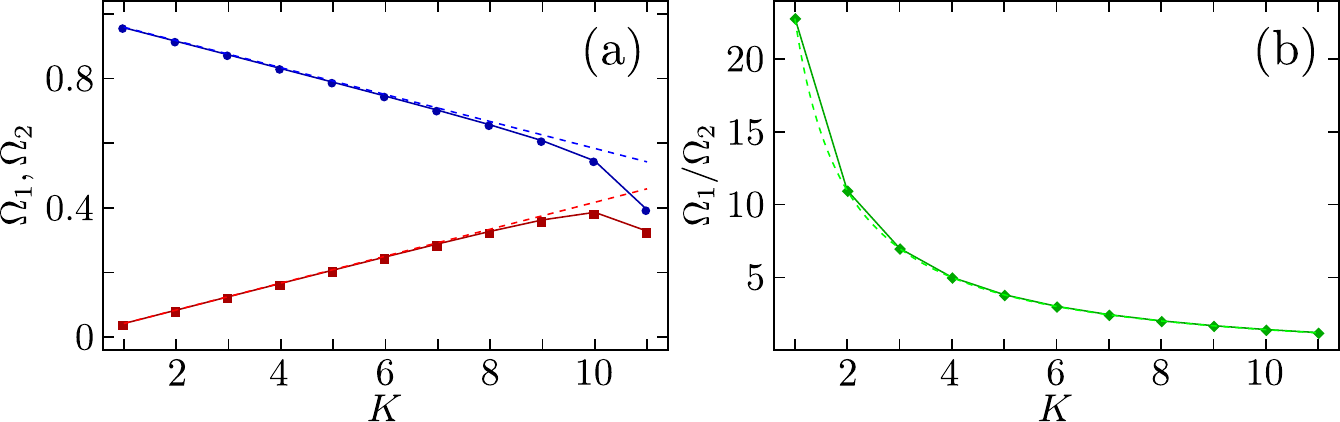}
\end{center}
  \caption{(a) Dependence of the average frequencies of elements of the small (blue) and large (red) clusters $\Omega_1$, $\Omega_2$ and (b) the ratio $\Omega_1/\Omega_2$ on the size of the small cluster $K$ for the corresponding two-cluster modes with a rotating phase difference $\theta_r\!\left(t\right)$ at $N=24$. Solid lines with markers were obtained numerically during direct numerical simulation of the system~\eqref{eq:main-system}; dotted lines -- using theoretical approximation Eq.~\eqref{eq:Omega_approx}. Parameters: $m=20$, $\alpha=1.6$, $\omega=1$.}
  \label{fig3}
\end{figure}

The Fig.~\ref{fig2} shows that the stability region of both clusters decreases with increasing $\beta$. To explain this effect, we performed experiments to calculate the average rotation frequencies of both clusters as a function of their sizes. The results are presented in Fig.~\ref{fig3} and Table 1. The rotation frequencies depend significantly on the cluster sizes. The elements of a small cluster rotate significantly (by times and even by orders of magnitude) faster compared to the elements of a large cluster. Thus, in the case of a significant difference in cluster sizes (small $\beta$), the influence of clusters on each other is not large. Clusters don't feel each other because the frequency resonances are absent and because of the weak influence of a small cluster on a large one. When $\beta$ tends to 1/2 (the sizes of clusters become close), the rotation frequencies in the clusters become close and the influence of clusters on each other becomes almost equal. A parametric resonance develops and one of the clusters becomes unstable.

It should be noted that at small $\beta=1/8$ (the large cluster is eight times larger than the small one), the stability region lies almost entirely in the region of the attractive coupling $\alpha<\pi/2$. As $\beta$ increases, the stability region of both clusters is also observed for the repulsive coupling $\alpha>\pi/2$. 

To verify the obtained analytical results, a series of computational experiments with the initial model~\eqref{eq:main-system} were performed.

\begin{figure}
\begin{center}
  \includegraphics[width=1\columnwidth]{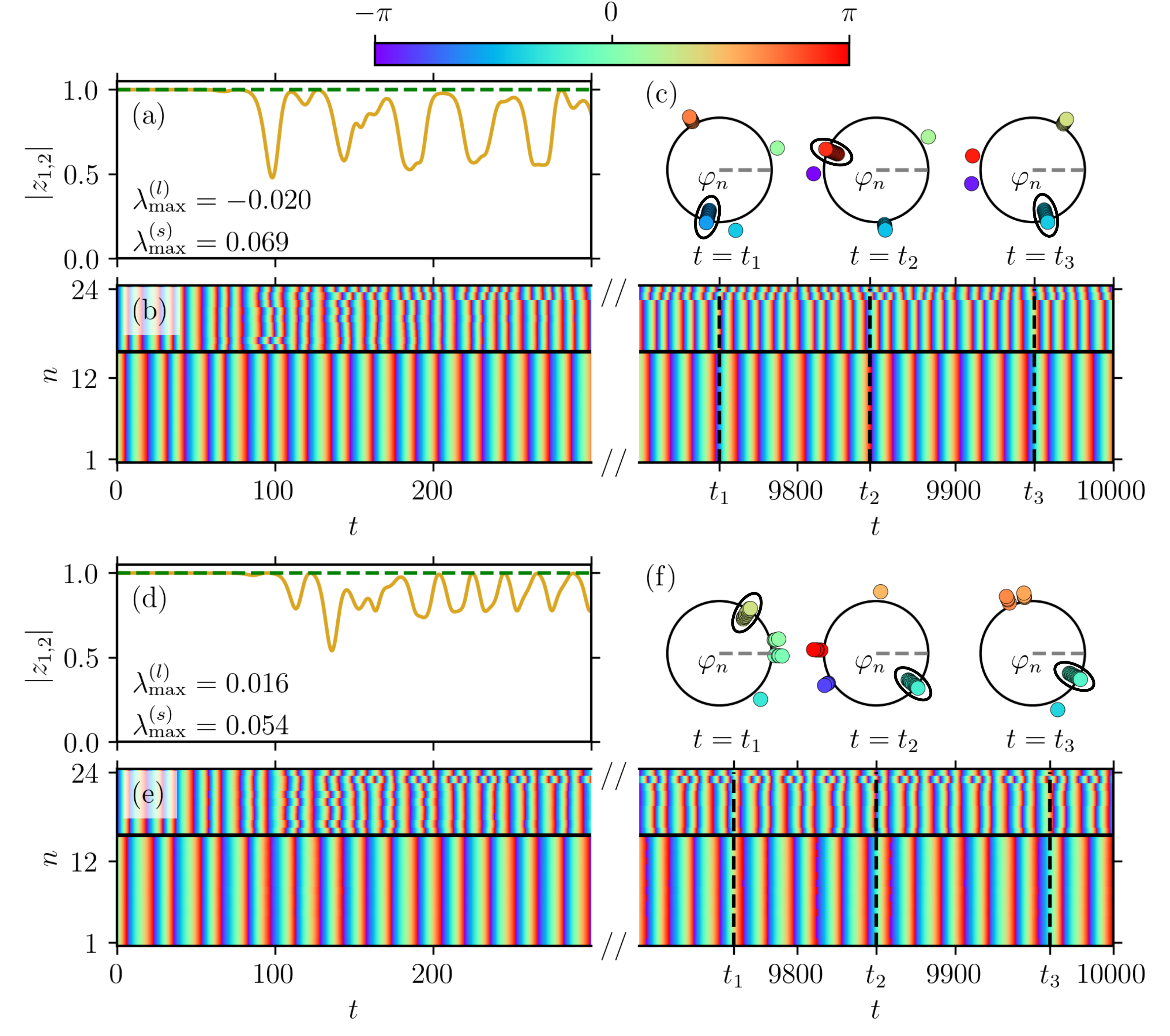}
\end{center}
  \caption{(a, d) Dynamics of the moduli of the clusters' global order parameters: small $|z_1|$ (hereafter highlighted in yellow) and large $|z_2|$ (hereafter highlighted in green). $\lambda^{(l)}_{\max}$ is the largest eigenvalue of the large cluster, $\lambda^{(s)}_{\max}$ is the largest eigenvalue of the small cluster. (b, e) Dynamics of the system~\eqref{eq:main-system} phases $\varphi_n$. (c, f) Phases $\varphi_n$ at fixed time moments $t=t_1$, $t=t_2$ and $t=t_3$. The black curve circles the elements of the largest stable cluster. (a, b, c) Point $(1)$ in Fig.~\ref{fig2}(d). Initial conditions: large cluster is stable, small cluster is unstable. The small cluster collapses, the large cluster remains stable. (d, e, f) Point $(2)$ in Fig.~\ref{fig2}(d). Initial conditions: large cluster is unstable, small cluster is unstable. The small cluster collapses, the large cluster stabilizes (its degree of instability is initially less). Parameters: $m=13.6$; (a,b,c) $\alpha = 1.25$; (d,e,f) $\alpha = 1.40$.}
\label{fig4}
\end{figure}

\begin{figure}
\begin{center}
  \includegraphics[width=1\columnwidth]{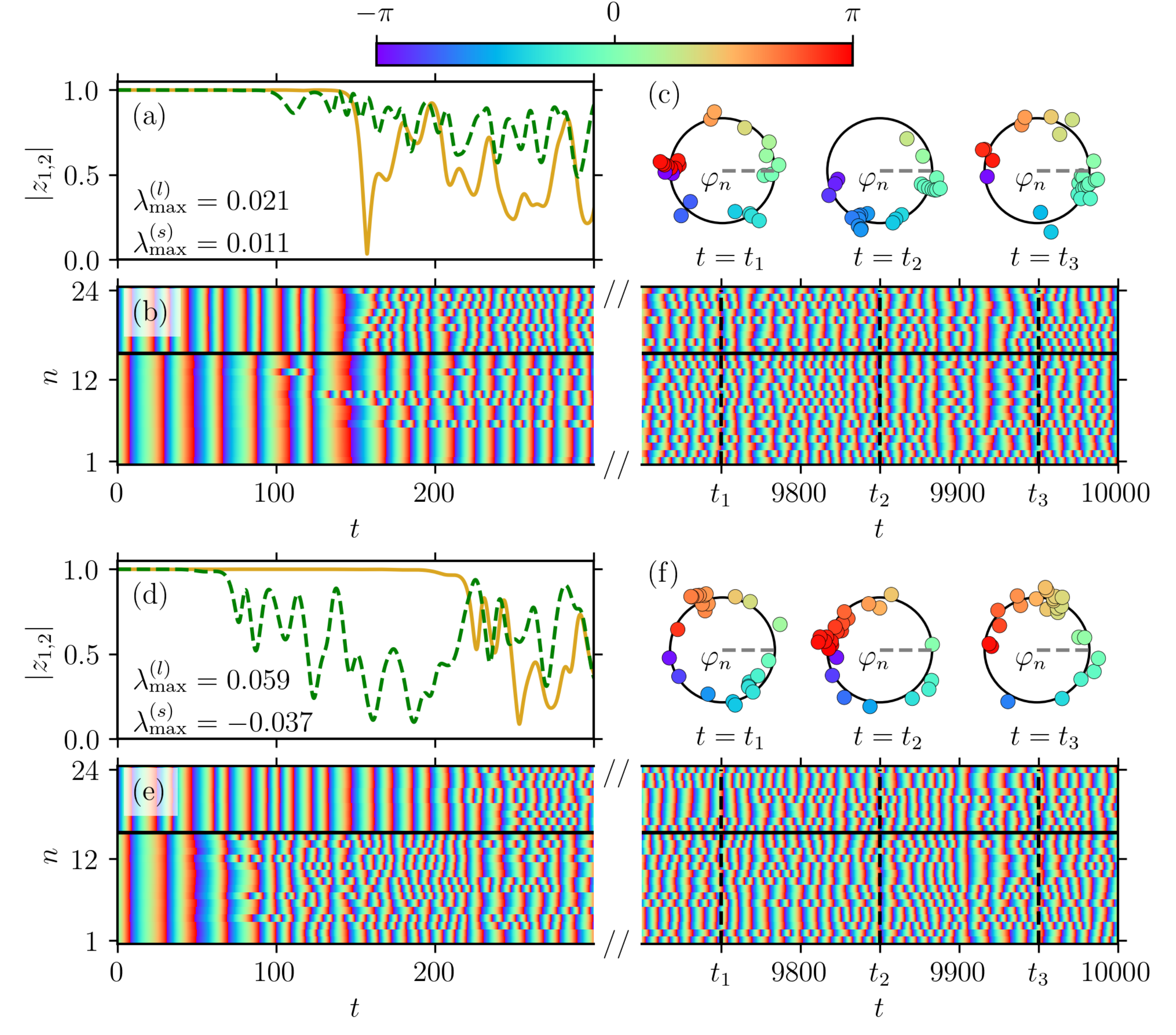}
\end{center}
  \caption{The figure shows the same as in Fig.~\ref{fig4}. (a, b, c) Point $(3)$ in Fig.~\ref{fig2}(d). Initial conditions: large cluster is unstable, small cluster is unstable. Both clusters collapse (the larger cluster collapses earlier due to the higher instability degree). (d, e, f) Point $(4)$ in Fig.~\ref{fig2}(d). Initial conditions: the large cluster is unstable, the small cluster is stable. The large cluster collapses, leading to instability of the small cluster. Parameters: $m=13.6$; (a,b,c) $\alpha = 1.77$; (d,e,f) $\alpha = 1.85$.}
\label{fig5}
\end{figure}

\begin{figure}
\begin{center}
  \includegraphics[width=1\columnwidth]{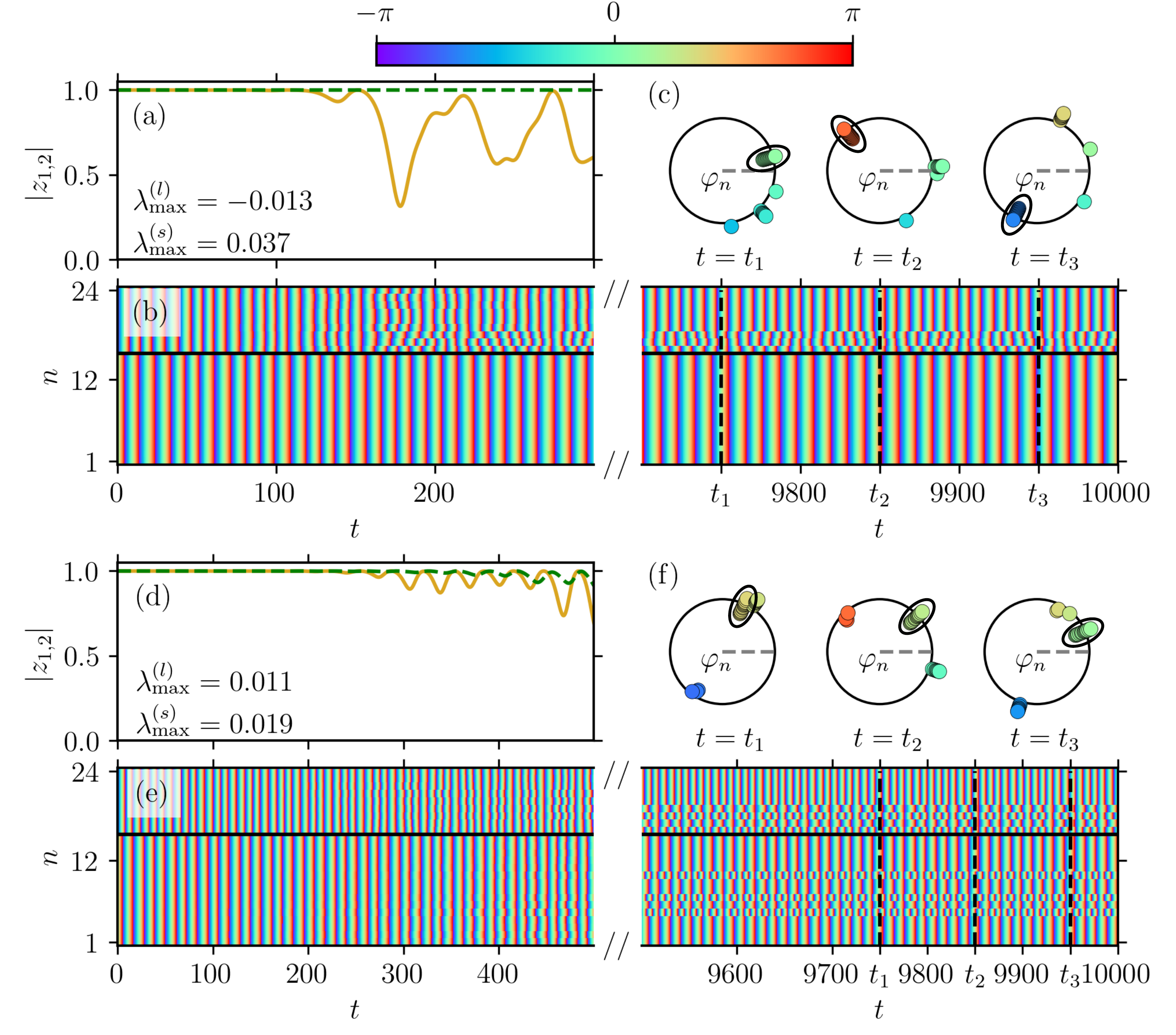}
\end{center}
  \caption{The figure shows the same as in Fig.~\ref{fig4}. (a, b, c) Point $(5)$ in Fig.~\ref{fig2}(d). Initial conditions: the large cluster is stable, the small one is unstable. The small cluster collapses, the large cluster remains stable. (d, e, f) Point $(6)$ in Fig.~\ref{fig2}(d). Initial conditions: the large and small clusters are unstable. Both clusters collapse (the small cluster collapses earlier due to the higher instability degree). Parameters: $m=40.0$; (a,b,c) $\alpha = 0.82$; (d,e,f) $\alpha = 0.96$.}
\label{fig6}
\end{figure}

\begin{figure}
\begin{center}
  \includegraphics[width=1\columnwidth]{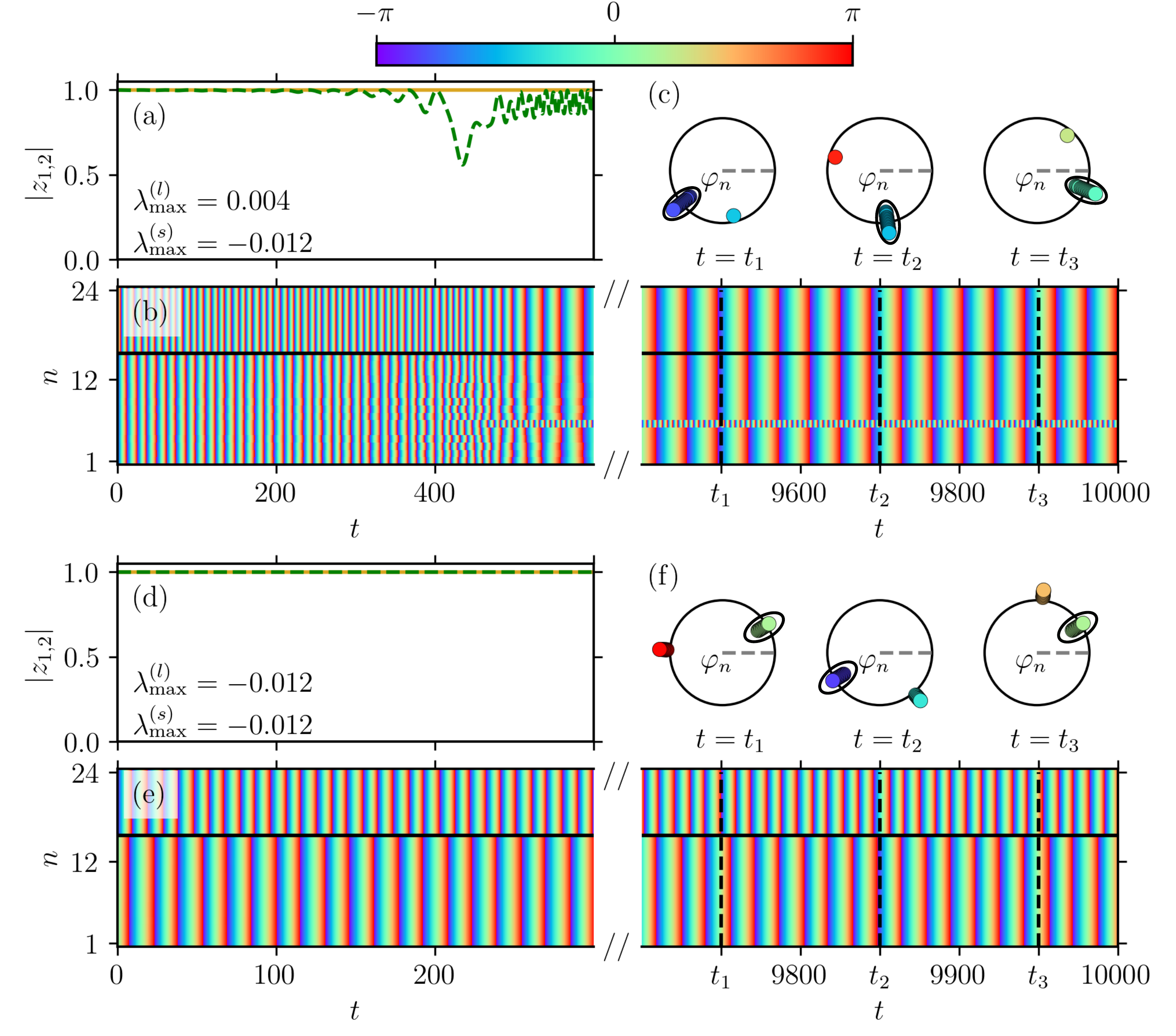}
\end{center}
  \caption{The figure shows the same as in Fig.~\ref{fig4}. (a, b, c) Point $(7)$ in Fig.~\ref{fig2}(d). Initial conditions: the large cluster is unstable, the small cluster is stable. The large cluster collapses: one element is isolated in it, and the rest of the oscillators join the small cluster; a solitary state is formed. (d, e, f) Point $(8)$ in Fig.~\ref{fig2}(d). Initial conditions: the large and small clusters are stable. Both clusters remain stable. Parameters: $m=40.0$; (a,b,c) $\alpha = 1.065$; (d,e,f) $\alpha = 1.36$.}
\label{fig7}
\end{figure}

\begin{figure}
\begin{center}
  \includegraphics[width=1\columnwidth]{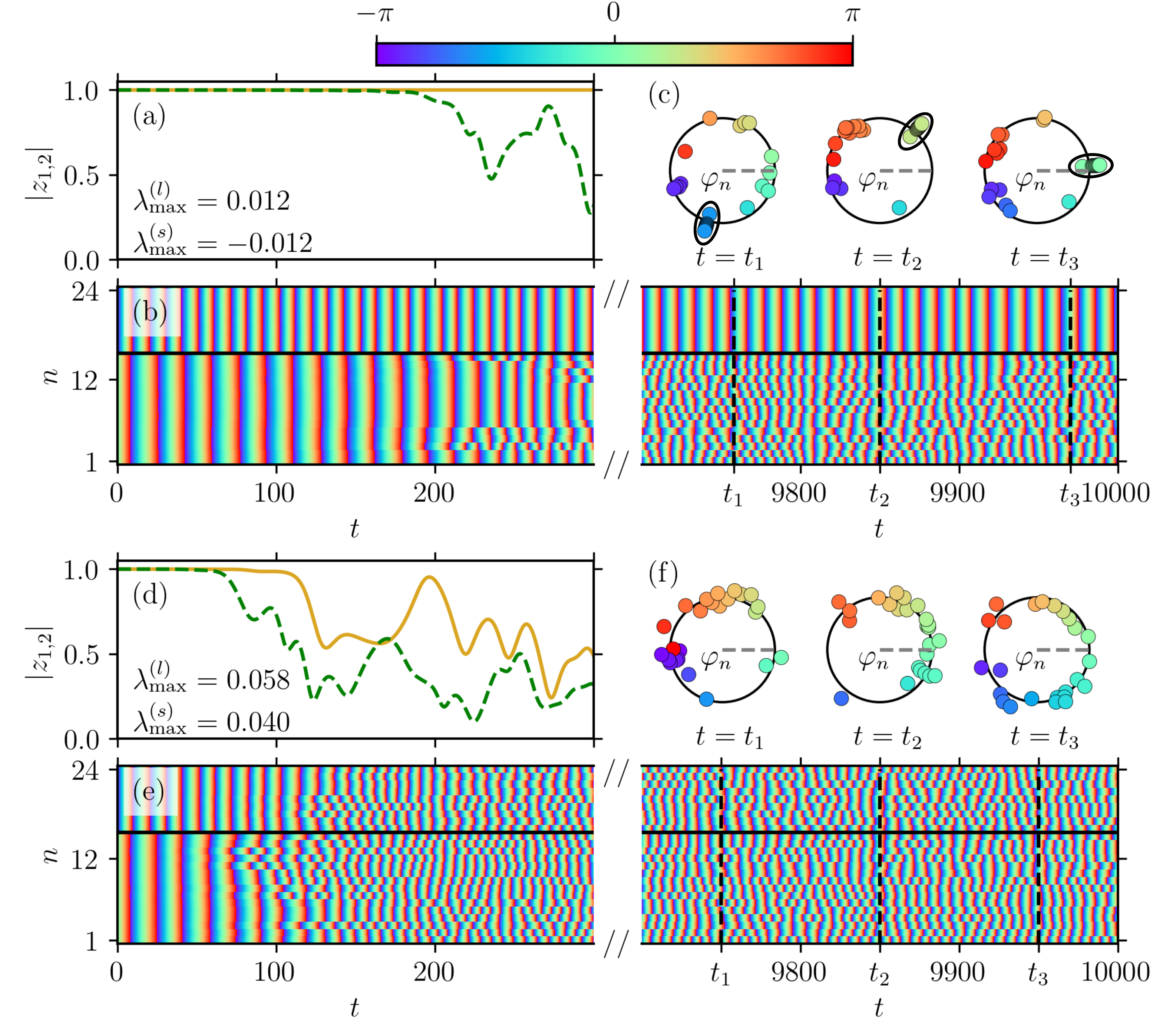}
\end{center}
  \caption{The figure shows the same as in Fig.~\ref{fig4}. (a, b, c) Point $(9)$ in Fig.~\ref{fig2}(d). Initial conditions: the large cluster is unstable, the small cluster is stable. The large cluster collapses, the small cluster remains stable, forming a chimera state. (d, e, f) Point $(10)$ in Fig.~\ref{fig2}(d). Initial conditions: the large and small clusters are unstable. Both clusters collapse (the large cluster collapses earlier due to the higher instability degree). Parameters: $m=40.0$; (a,b,c) $\alpha = 1.67$; (d,e,f) $\alpha = 1.93$.}
\label{fig8}
\end{figure}

Figs.~\ref{fig4}--\ref{fig8} show time diagrams (panels~(b) and (e)) of the phase dynamics $\varphi_n$ of the system~\eqref{eq:main-system} with $N = 24$ elements, $\beta=3/8$ (i.e. small cluster has size $K=9$, large -- $M=N-K=15$) for two fixed inertia values $m=13.6$ and $m=40.0$ and different values of the parameter $\alpha$ taken from all colored regions on the map in Fig.~\ref{fig2}(d). The time diagrams are supplemented with plots (panels (a) and (d)) of the evolution of the global order parameters $|z_1|$ for a small cluster and $|z_2|$ for a large cluster, computed according to the definitions
\begin{equation}
  z_1=\frac{1}{K}\sum_{{\tilde{n}}=1}^{K}e^{i\varphi_{\tilde{n}}},\quad
  z_2=\frac{1}{N-K}\sum_{{\tilde{n}}=K+1}^{N}e^{i\varphi_{\tilde{n}}},
\end{equation}
and maximum values of eigenvalues determining the clusters' stability: $\lambda^{(l)}_{\max}$ -- the largest eigenvalue of a large cluster, $\lambda^{(s)}_{\max}$ -- the largest eigenvalue of a small cluster (on panels (a) and (d)). The figures also represent the values of the phases $\varphi_n$ at fixed time moments $t=t_1$, $t=t_2$ and $t=t_3$ (panels (c) and (f)). Note that in the case of the in-phase mode, the modulus of the global order parameter is equal to one. The numerical simulation of the system~\eqref{eq:main-system} was performed with initial conditions close to the realization of the two-cluster rotatory regime, regardless of the stability of both clusters.

\begin{figure}
\begin{center}
  \includegraphics[width=1\columnwidth]{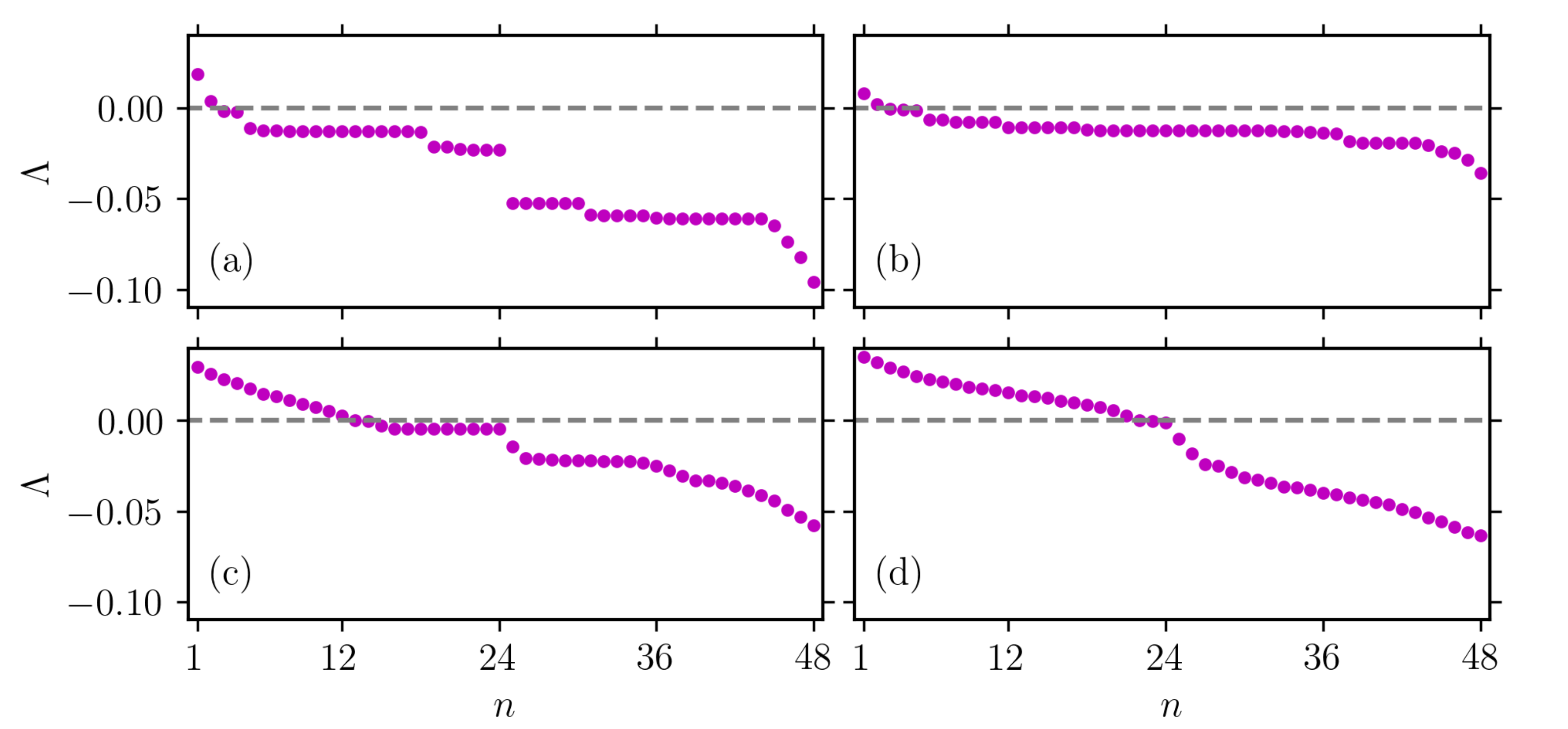}
\end{center}
  \caption{The spectrum of the Lyapunov exponents $\Lambda_n$ ($n=1,2,\dots, 2N$) for the regimes observed at (a) $m=13.6$, $\alpha=1.25$ (see Fig.~\ref{fig4}(a,b,c)); (b) $m=40.0$, $\alpha=0.96$ (see Fig.~\ref{fig6}(d,e,f)); (c) $m=40.0$, $\alpha=1.67$ (see Fig.~\ref{fig8}(a,b,c)); (d) $m=40.0$, $\alpha=1.93$ (see Fig.~\ref{fig8}(d,e,f)).}
\label{fig9}
\end{figure}

\textit{Case 1}: $m=13.6$, $\alpha=1.25$ (see Fig.~\ref{fig4}(a,b,c). Point (1) from the green area in Fig.~\ref{fig2}(d)). 
Initial conditions: the large cluster is stable, the small cluster is unstable. The small cluster is destroyed, the large cluster is preserved. In place of the small cluster, a cluster of in-phase rotating elements of smaller size ($7$ elements) is formed, and two elements rotate with different average frequencies from the others. A weak chaos is observed in the ensemble due to these elements (see Fig.~\ref{fig9}(a), where only two Lyapunov exponents are positive).

\textit{Case 2}: $m=13.6$, $\alpha=1.40$ (see Fig.~\ref{fig4}(d,e,f). Point (2) from the red area in Fig.~\ref{fig2}(d)).
Initial conditions: the large and small clusters are unstable. In this case the unexpected situation occurs. The small cluster collapses, but the large cluster is stabilized due to the impact from the elements of the small cluster due to the low instability degree. At the same time, the small cluster breaks up as in the previous case into a smaller cluster ($7$ elements) and two independently rotating elements.

\textit{Case 3}: $m=13.6$, $\alpha=1.77$ (see Fig.~\ref{fig5}(a,d,c). Point (3) from the red area in Fig.~\ref{fig2}(d)).
Initial conditions: the large and small clusters are unstable. Both clusters collapse. At the same time, the large cluster collapses earlier due to the greater instability degree. Chaotic behavior is observed in the ensemble. The dynamics becomes more complex through the sequential separation of individual elements from the clusters.

\textit{Case 4}: $m=13.6$, $\alpha=1.85$ (see Fig.~\ref{fig5}(d,e,f). Point (4) from the yellow area in Fig.~\ref{fig2}(d)).
Initial conditions: the large cluster is unstable, the small cluster is stable. The large cluster collapses, causing the small cluster to become unstable. At the same time, the small cluster lives long enough. As a result, chaotic behavior takes place in the ensemble. As in the previous case, there is a complication of the dynamics through the sequential separation of individual elements from the clusters. At the same time, on spatio-temporal diagrams one can see areas of coherent behavior of oscillators.

\textit{Case 5}: $m=40.0$, $\alpha=0.82$ (see Fig.~\ref{fig6}(a,b,c). Point (5) from the green area in Fig.~\ref{fig2}(d)).
Initial conditions: the large cluster is stable, the small cluster is unstable. The small cluster collapses, the large cluster remains stable. In the small cluster, the transition to nonsynchronous behavior (at $t\approx 180$) occurs rather sharply (the parameter of order $z_1$ decreases sharply). However, desynchronization in the small cluster has no effect on synchronization in the large cluster. An in-phase cluster of $6$ elements appears in place of the small cluster.

\textit{Case 6}: $m=40.0$, $\alpha=0.96$ (see Fig.~\ref{fig6}(d,e,f). Point (6) from the red area in Fig.~\ref{fig2}(d)).
Initial conditions: the large and small clusters are unstable. Both clusters collapse. The small cluster collapses earlier than the large one due to a higher instability degree. Destruction occurs through the appearance of individual oscillators with different frequencies. There is weak chaos (Fig.~\ref{fig9}(b) with two Lyapunov exponents are positive).

\textit{Case 7}: $m=40.0$, $\alpha=1.065$ (see Fig.~\ref{fig7}(a,b,c). Point (7) from the yellow area in Fig.~\ref{fig2}(d)).
Initial conditions: the large cluster is unstable, the small cluster is stable. The large cluster collapses through the appearance of a solitary state. The remaining oscillators of the large cluster are captured by the small cluster.

\textit{Case 8}: $m=40.0$, $\alpha=1.36$ (see Fig.~\ref{fig7}(d,e,f). Point (8) from the blue area in Fig.~\ref{fig2}(d)).
Initial conditions: the large and small clusters are stable.
Both clusters remain stable. In both clusters in-phase rotational modes with different rationally incommensurable frequencies are realized (see Table 1, where the frequency ratio $\Omega_1/\Omega_2 \approx 1.6756$).
According to Table 1, full synchronization (or near synchronization) of rotations in clusters is possible under certain conditions. For $N=24$ elements, it occurs for small cluster sizes $K$ equal to (i) $4$ -- synchronization close to the ratio $5:1$; (ii) $6$ -- synchronization close to the ratio $3:1$, (iii) synchronization close to the ratio $2:1$. Note that using the expressions~\eqref{eq:Omega}, we can determine the values of the control parameters of the system~\eqref{eq:main-system} at which the frequency ratio $\Omega_1/\Omega_2$ is a given rational number $p/q$, i.e., there is a frequency synchronization $p:q$.

\textit{Case 9}: $m=40.0$, $\alpha=1.67$ (see Fig.~\ref{fig8}(a,b,c). Point (9) from the yellow area in Fig.~\ref{fig2}(d)). 
Initial conditions: the large cluster is unstable, the small cluster is stable. The large cluster collapses and its elements become chaotic (chaos is localized in space), the small cluster remains stable. Thus, a chimera state is formed. The behavior in the region of nonsynchronous oscillations is hyperchaotic: 12 Lyapunov exponents are positive (see Fig.~\ref{fig9}(c)).

\textit{Case 10}: $m=40.0$, $\alpha=1.93$ (see Fig.~\ref{fig8}(d,e,f). Point (10) from the red area in Fig.~\ref{fig2}(d)). 
Initial conditions: the large cluster is unstable, the small cluster is unstable. Both clusters collapse. The larger cluster collapses earlier due to the higher instability degree. A global (in the whole ensemble) hyperchaotic regime emerges. The $20$ Lyapunov exponents are positive (Fig.~\ref{fig9}(d)).

The results of the experiments with the original system~\eqref{eq:main-system} are in perfect agreement with the analytical results obtained with the reduced system and presented in the stability maps of the two-cluster rotational state (see Fig.~\eqref{fig2}). 

\begin{figure}
\begin{center}
  \includegraphics[width=1\columnwidth]{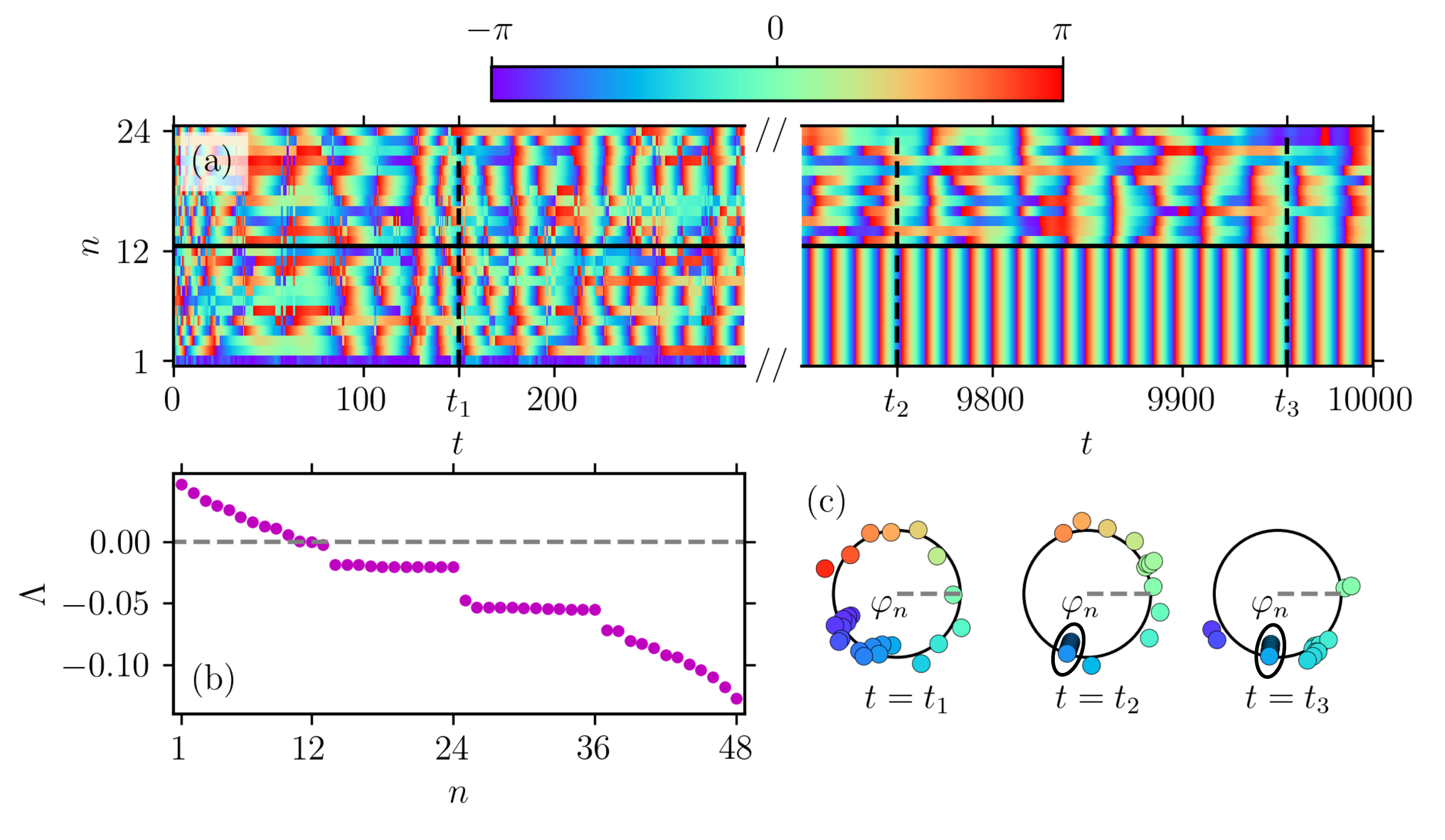}
\end{center}
  \caption{The chimera state emergence. (a) Dynamics of the system~\eqref{eq:main-system} phases $\varphi_n$. (b) The spectrum of the Lyapunov exponents $\Lambda_n$. (c) Phases $\varphi_n$ at fixed times $t=t_1$, $t=t_2$, and $t=t_3$. The black curve circles the elements of the synchronous cluster. Initial conditions: the phases $\varphi_n(0)$ are uniformly distributed on the interval $[-\pi, \pi]$, $\dot{\varphi}_n(0) = 0.0$. Parameters: $m=13.6$, $\alpha=1.65$.}
\label{fig10}
\end{figure}

All the above experiments were performed with initial conditions close to the realization of the two-cluster rotatory regime. It is shown (see Fig.~\ref{fig1}) that the system under consideration has the property of multistability and can, as a consequence, demonstrate a wide range of different dynamical modes. We especially note the case when chimera states are realized in the system. Fig.~\ref{fig10} shows the space-time diagram, the spectrum of Lyapunov exponents and the oscillators' phases $\varphi_n$ distribution for an ensemble of $N=24$ elements at $m=13.6$, $\alpha=1.65$. During the evolution, the ensemble has split into two equal parts. One half exhibits in-phase synchronous behavior. In the other half, a nonsynchronous hyperchaotic mode is established, i.e., a chimera state takes place. This state is dominant under random uniformly distributed initial conditions (phases and instantaneous frequencies).

\section{Conclusion}
\label{sec:con}
In this paper, the existence and stability conditions of two-cluster rotatory regimes in an ensemble of globally coupled phase oscillators with inertia were analyzed. The question of two-cluster rotatory regimes existence depending on two main control parameters, effective inertia and phase lag (Sakaguchi parameter), was studied. In particular, an equation describing the boundaries of the existence regions of the rotatory regime of interest was derived (see Eq.~\eqref{eq:eroi}). It was shown that two types of two-cluster rotatory regimes can exist. The first one is characterized by a constant intercluster phase difference, the second one is characterized by a periodic rotatory phase difference.

It has been shown that two-cluster rotatory regime with constant phase difference is unstable at any values of the control parameters. The exception is the case when the system consists of an even number of elements. In this situation, in the repulsive interaction region, one of the typical regimes is the antiphase state, when each of the two clusters has an equal number of oscillators, and the phase difference between them is $\pi/2$. In turn, there are no two-cluster regimes with periodically varying (taking into account the multiplicity $2\pi$) phase difference with the same number of particles in the separated coherent subgroups. For this kind of states it is always possible to distinguish a large and a small cluster. In particular, the solitary state belongs to this class of regimes.

It has been shown that the two-cluster regime, in which one in-phase subset rotates relative to the other, collapses due to the loss of stability of one of the clusters (small or large). At the same time, the other cluster remains stable in the linear approximation and can in some cases be preserved as a whole, but can disintegrate under the influence of strong fluctuations caused by the desynchronized part of the system. Destruction of the two-cluster regime due to simultaneous loss of stability by both clusters is also possible, but less typical. As a result, the following results were obtained:
\begin{enumerate}
    \item There is no stable two-cluster periodic rotational regimes \textit{with constant phase difference} of oscillators in clusters, besides the anti-phase mode in the repulsive interaction region.
    \item There is no stable two-cluster periodic rotational regimes \textit{with the same number of elements} in both clusters with periodically varying phase difference.
    \item The number of two-cluster periodic rotational modes in the ensemble of $N$ elements is $\lfloor(N-3)/2\rfloor$ (floor function).
    \item Given an ensemble size $N$ and a given ratio $\beta=K/N$ of the small cluster size $K$ to the ensemble size $N$, there are four types of parameter regions in which different regimes are realized:
    
    a) both clusters are stable;

b) large cluster is stable, small cluster is unstable; 

c) small cluster is stable, large cluster is unstable;

d) both clusters are unstable.
\end{enumerate}

In all of the cases cited, it is possible in place of the unstable cluster(s): 
\begin{itemize}
    \item emergence of a two-cluster structure with a different partitioning in the number of elements between clusters;
    \item appearance of three-, four-, and other cluster structures of both regular and chaotic rotations;
    \item simultaneous formation of a regular cluster and a group of oscillators with chaotic rotation (e.g., chimera) or regimes characterized by chaotic rotation of all elements.
\end{itemize}

The results were validated by direct numerical simulations.
The described mechanisms of two-cluster regimes destruction may be interesting from the point of view of understanding the scenarios of emergence of complex multicluster states with regular and chaotic behavior of individual elements in ensembles containing a large number of elements.

The authors thank D.~S.~Khorkin for the technical advice. This work was supported by the Ministry of Science and Higher Education of the Russian Federation (Secs.~2, 3, project No.~0729-2020-0036), the Russian Science Foundation (Sec.~4, project No.~22-12-00348 and Sec.~5, project No.~23-12-00180).
\clearpage

\begin{table}[h!]
\begin{center}
  \begin{tabular}{|p{1.0cm}||p{1.0cm}|p{1.0cm}|p{1.0cm}|p{1.0cm}|p{0.9cm}|p{0.9cm}|p{0.9cm}|p{0.9cm}|p{0.9cm}|p{0.9cm}|p{0.9cm}|}
  \hline
    $K$ & 1 & 2 & 3 & 4 & 5 & 6 & 7 & 8 & 9 & 10 & 11 \\
  \hline\hline
    $\Omega_1/\Omega_2$ & 22.8127 & 10.9681 & 6.9931 & 5.0007 & 3.8037 & 3.0053 & 2.4350 & 2.0074 & 1.6756 & 1.4117 & 1.2011 \\
  \hline
  \end{tabular}
  \caption{Dependence of the ratio $\Omega_1/\Omega_2$ on the size of the small cluster $K$ for the corresponding two-cluster modes with periodic intercluster phase difference, obtained numerically during direct numerical simulation of the system~\eqref{eq:main-system}. Parameters: $N=24$, $m=20$, $\alpha=1.6$, $\omega=1$.}
\end{center}
\end{table}

\clearpage

 \bibliographystyle{elsarticle-num} 
 \bibliography{cas-refs}

\begin{thebibliography}{10}
\expandafter\ifx\csname url\endcsname\relax
  \def\url#1{\texttt{#1}}\fi
\expandafter\ifx\csname urlprefix\endcsname\relax\def\urlprefix{URL }\fi
\expandafter\ifx\csname href\endcsname\relax
  \def\href#1#2{#2} \def\path#1{#1}\fi

\bibitem{hoppensteadt2012}
F.~C. Hoppensteadt, E.~M. Izhikevich, Weakly Connected Neural Networks, Vol.
  126 of Applied Mathematical Sciences, Springer Science \& Business Media,
  2012.
\newblock \href {https://doi.org/10.1007/978-1-4612-1828-9}
  {\path{doi:10.1007/978-1-4612-1828-9}}.

\bibitem{tinsley2012}
M.~R. Tinsley, S.~Nkomo, K.~Showalter, Chimera and phase-cluster states in
  populations of coupled chemical oscillators, Nature Physics 8~(9) (2012)
  662--665.
\newblock \href {https://doi.org/10.1038/nphys2371}
  {\path{doi:10.1038/nphys2371}}.

\bibitem{ding2019}
J.~Ding, I.~Belykh, A.~Marandi, M.-A. Miri, Dispersive versus dissipative
  coupling for frequency synchronization in lasers, Physical Review Applied
  12~(5) (2019) 054039.
\newblock \href {https://doi.org/10.1103/physrevapplied.12.054039}
  {\path{doi:10.1103/physrevapplied.12.054039}}.

\bibitem{dorfler2013}
F.~D\"{o}rfler, M.~Chertkov, F.~Bullo, Synchronization in complex oscillator
  networks and smart grids, Proceedings of the National Academy of Sciences
  110~(6) (2013) 2005--2010.
\newblock \href {https://doi.org/10.1073/pnas.1212134110}
  {\path{doi:10.1073/pnas.1212134110}}.

\bibitem{acebron2005}
J.~A. Acebr{\'{o}}n, L.~L. Bonilla, C.~J.~P. Vicente, F.~Ritort, R.~Spigler,
  The kuramoto model: A simple paradigm for synchronization phenomena, Reviews
  of Modern Physics 77~(1) (2005) 137--185.
\newblock \href {https://doi.org/10.1103/revmodphys.77.137}
  {\path{doi:10.1103/revmodphys.77.137}}.

\bibitem{barreto2008}
E.~Barreto, B.~Hunt, E.~Ott, P.~So, Synchronization in networks of networks:
  The onset of coherent collective behavior in systems of interacting
  populations of heterogeneous oscillators, Physical Review E 77~(3) (2008)
  036107.
\newblock \href {https://doi.org/10.1103/physreve.77.036107}
  {\path{doi:10.1103/physreve.77.036107}}.

\bibitem{ott2008}
E.~Ott, T.~M. Antonsen, Low dimensional behavior of large systems of globally
  coupled oscillators, Chaos: An Interdisciplinary Journal of Nonlinear Science
  18~(3) (2008) 037113.
\newblock \href {https://doi.org/10.1063/1.2930766}
  {\path{doi:10.1063/1.2930766}}.

\bibitem{hong2007}
H.~Hong, H.~Chat{\'{e}}, H.~Park, L.-H. Tang, Entrainment transition in
  populations of random frequency oscillators, Physical Review Letters 99~(18)
  (2007) 184101.
\newblock \href {https://doi.org/10.1103/physrevlett.99.184101}
  {\path{doi:10.1103/physrevlett.99.184101}}.

\bibitem{pikovsky2008}
A.~Pikovsky, M.~Rosenblum, Partially integrable dynamics of hierarchical
  populations of coupled oscillators, Physical Review Letters 101~(26) (2008)
  264103.
\newblock \href {https://doi.org/10.1103/physrevlett.101.264103}
  {\path{doi:10.1103/physrevlett.101.264103}}.

\bibitem{maistrenko2004}
Y.~Maistrenko, O.~Popovych, O.~Burylko, P.~A. Tass, Mechanism of
  desynchronization in the finite-dimensional kuramoto model, Physical Review
  Letters 93~(8) (2004) 084102.
\newblock \href {https://doi.org/10.1103/physrevlett.93.084102}
  {\path{doi:10.1103/physrevlett.93.084102}}.

\bibitem{dorfler2011}
F.~D\"{o}rfler, F.~Bullo, On the critical coupling for kuramoto oscillators,
  {SIAM} Journal on Applied Dynamical Systems 10~(3) (2011) 1070--1099.
\newblock \href {https://doi.org/10.1137/10081530x}
  {\path{doi:10.1137/10081530x}}.

\bibitem{martens2009}
E.~A. Martens, E.~Barreto, S.~H. Strogatz, E.~Ott, P.~So, T.~M. Antonsen, Exact
  results for the kuramoto model with a bimodal frequency distribution,
  Physical Review E 79~(2) (2009) 026204.
\newblock \href {https://doi.org/10.1103/physreve.79.026204}
  {\path{doi:10.1103/physreve.79.026204}}.

\bibitem{laing2009}
C.~R. Laing, The dynamics of chimera states in heterogeneous kuramoto networks,
  Physica D: Nonlinear Phenomena 238~(16) (2009) 1569--1588.
\newblock \href {https://doi.org/10.1016/j.physd.2009.04.012}
  {\path{doi:10.1016/j.physd.2009.04.012}}.

\bibitem{ermentrout1991}
B.~Ermentrout, An adaptive model for synchrony in the firefly pteroptyx
  malaccae, Journal of Mathematical Biology 29~(6) (1991) 571--585.
\newblock \href {https://doi.org/10.1007/bf00164052}
  {\path{doi:10.1007/bf00164052}}.

\bibitem{tumash2019}
L.~Tumash, S.~Olmi, E.~Sch\"{o}ll, Stability and control of power grids with
  diluted network topology, Chaos: An Interdisciplinary Journal of Nonlinear
  Science 29~(12) (2019) 123105.
\newblock \href {https://doi.org/10.1063/1.5111686}
  {\path{doi:10.1063/1.5111686}}.

\bibitem{brister2020}
B.~N. Brister, V.~N. Belykh, I.~V. Belykh, When three is a crowd: Chaos from
  clusters of kuramoto oscillators with inertia, Physical Review E 101~(6)
  (2020) 062206.
\newblock \href {https://doi.org/10.1103/physreve.101.062206}
  {\path{doi:10.1103/physreve.101.062206}}.

\bibitem{olmi2015}
S.~Olmi, Chimera states in coupled kuramoto oscillators with inertia, Chaos: An
  Interdisciplinary Journal of Nonlinear Science 25~(12) (2015) 123125.
\newblock \href {https://doi.org/10.1063/1.4938734}
  {\path{doi:10.1063/1.4938734}}.

\bibitem{maistrenko2017}
Y.~Maistrenko, S.~Brezetsky, P.~Jaros, R.~Levchenko, T.~Kapitaniak, Smallest
  chimera states, Physical Review E 95~(1) (2017) 010203.
\newblock \href {https://doi.org/10.1103/physreve.95.010203}
  {\path{doi:10.1103/physreve.95.010203}}.

\bibitem{medvedev2021}
G.~S. Medvedev, M.~S. Mizuhara, Stability of clusters in the second-order
  kuramoto model on random graphs, Journal of Statistical Physics 182~(2)
  (2021) 1--22.
\newblock \href {https://doi.org/10.1007/s10955-021-02708-2}
  {\path{doi:10.1007/s10955-021-02708-2}}.

\bibitem{jaros2015}
P.~Jaros, Y.~Maistrenko, T.~Kapitaniak, Chimera states on the route from
  coherence to rotating waves, Physical Review E 91~(2) (2015) 022907.
\newblock \href {https://doi.org/10.1103/physreve.91.022907}
  {\path{doi:10.1103/physreve.91.022907}}.

\bibitem{jaros2018}
P.~Jaros, S.~Brezetsky, R.~Levchenko, D.~Dudkowski, T.~Kapitaniak,
  Y.~Maistrenko, Solitary states for coupled oscillators with inertia, Chaos:
  An Interdisciplinary Journal of Nonlinear Science 28~(1) (2018) 011103.
\newblock \href {https://doi.org/10.1063/1.5019792}
  {\path{doi:10.1063/1.5019792}}.

\bibitem{munyayev2022}
V.~O. Munyayev, M.~I. Bolotov, L.~A. Smirnov, G.~V. Osipov, I.~V. Belykh,
  Stability of rotatory solitary states in kuramoto networks with inertia,
  Physical Review E 105~(2) (2022) 024203.
\newblock \href {https://doi.org/10.1103/physreve.105.024203}
  {\path{doi:10.1103/physreve.105.024203}}.

\bibitem{munyayev2023}
V.~O. Munyayev, M.~I. Bolotov, L.~A. Smirnov, G.~V. Osipov, I.~Belykh, Cyclops
  states in repulsive kuramoto networks: The role of higher-order coupling,
  Phys. Rev. Lett. 130 (2023) 107201.
\newblock \href {https://doi.org/10.1103/physrevlett.130.107201}
  {\path{doi:10.1103/physrevlett.130.107201}}.

\bibitem{belykh2016}
I.~V. Belykh, B.~N. Brister, V.~N. Belykh, Bistability of patterns of synchrony
  in kuramoto oscillators with inertia, Chaos: An Interdisciplinary Journal of
  Nonlinear Science 26~(9) (2016) 094822.
\newblock \href {https://doi.org/10.1063/1.4961435}
  {\path{doi:10.1063/1.4961435}}.

\bibitem{sakaguchi2006}
H.~Sakaguchi, Instability of synchronized motion in nonlocally coupled neural
  oscillators, Physical Review E 73~(3) (2006) 031907.
\newblock \href {https://doi.org/10.1103/physreve.73.031907}
  {\path{doi:10.1103/physreve.73.031907}}.

\bibitem{andronov2013}
A.~Andronov, A.~Vitt, S.~Khaikin, Theory of Oscillators: Adiwes International
  Series in Physics, Vol.~4 of International series of monographs in physics,
  Elsevier Science, 2013.

\end{thebibliography}





% \end{thebibliography}
\end{document}